\documentclass[10pt,letterpaper]{article}
\usepackage{opex3,cite}

\usepackage{amssymb,amsmath,placeins}

\newcommand{\EQE}{\text{EQE}_\text{sc}}

\newcommand{\dd}{\partial}
\newcommand{\rd}{\mathrm{d}}
\newcommand{\pd}[2]{\frac{\partial #1}{\partial #2}}
\newcommand{\td}[2]{\frac{\rd #1}{\rd #2}}

\newcommand{\ve}[1]{\mathbf{#1}}
\newcommand{\ez}{\ve{\hat z}}
\newcommand{\ex}{\ve{\hat x}}

\newcommand{\eps}{\epsilon}
\newcommand{\bs}[1]{\boldsymbol{#1}}

\newcommand{\beq}{\begin{equation}}
\newcommand{\eeq}{\end{equation}}

\newcommand{\re}{\mbox{Re}}
\newcommand{\im}{\mbox{Im}}

\newcommand{\curl}{\bs{\nabla} \times}
\newcommand{\Div}{\bs{\nabla} \cdot}
\newcommand{\grad}{\bs{\nabla}}

\begin{document}

\title{Microcavity effects on the generation, fluorescence, and diffusion of excitons in organic solar cells}

\author{G. Kozyreff$^1$, D. C. Urbanek$^2$, L.T. Vuong$^3$, O. Nieto Silleras$^1$, and J.~Martorell$^{2,4}$}

\address{$^1$Universit\'e libre de Bruxelles (ULB), C.P. 231,  Campus de la  Plaine, B-1050 Bruxelles, Belgium\\
$^2$ICFO-Institut de Ciencies Fotoniques, Mediterranean Technology Park, 08860 Castelldefels (Barcelona), Spain\\ 
$^3$ Physics Department, City University of New York- Queens College and the Graduate Center, Flushing, NY, USA ,\\
$^4$Departament de F\'isica i Enginyeria Nuclear, Universitat Polit\`ecnica de Catalunya, Terassa, Spain}

\email{gkozyref@ulb.ac.be} 



\begin{abstract}
We compute the short-circuit diffusion current of excitons in an organic solar cell, with special emphasis on fluorescence losses. The exciton diffusion length is not uniform but varies with its position within the device, even with moderate fluorescence quantum efficiency. With large quantum efficiencies, the rate of fluorescence can be strongly reduced with proper choices of the geometrical and dielectric parameters. In this way, the diffusion length can be increased and the device performance significantly improved.
\end{abstract}






\section{Introduction}

A fundamental obstacle to the efficient conversion of solar radiation into electric current by organic material is the short diffusion length of photo-generated excitons. These need to survive long enough to reach a dissociation site, where they can separate into holes and electrons. Hence the thickness of the donor and acceptor layers in bilayer hetero-junctions is severely limited. That problem is overcome in bulk hetero-junctions, but it is then replaced by the fact that electrons and holes need to percolate through an intricate structure. They are thus less mobile and  can be trapped in dead ends \cite{Forrest-2005}. The aim of this paper is to show that the exciton diffusion length can significantly be modified by the micro-cavity formed by the various layers in an organic solar cell, and that a proper understanding of this effect could lead to a better performance of bilayer hetero-junctions. 

Most exciton transport models boil down to the diffusion equation
\begin{align}
0&=L^2\td{^2\rho}{z^2}-\rho +g(z),&
L^2&=D\tau,
\label{Diff1}
\end{align}
in the photoactive material, where $\rho$ is the exciton concentration, $D$ is the diffusion constant, $\tau$ is the lifetime, and $g(z)$ is the source term. Different modeling considerations lead to different forms of $g(z)$ and different boundary conditions \cite{DeVore-1956,Ghosh-1978,Desormeaux-1993}. A comparison between the main models concluded that they all qualitatively reproduce the photovoltaic response of a given organic solar cell but that none was fully accurate \cite{Harrison-1997}. More recently, (\ref{Diff1}) was used to experimentally determine the diffusion length $L$ from the optical response of solar cells~\cite{Stubinger-2001,Lunt-2009,Bergemann-2011}. 

However, if fluorescence is taken into account, (\ref{Diff1}) should be modified as
\begin{align}
0&=L^2\td{^2\rho}{z^2}-b(z)\rho +g(z),&
L^2&=D\tau,
\label{Diff2}
\end{align}
where the factor $b(z)$ is the decay rate normalized to the bulk value and is now space-dependent. Thus, the diffusion length is locally corrected by the factor $1/\sqrt{b(z)}$. The function $b(z)$ is determined by the optical thicknesses of the layers that make up the solar cell and the wavelength of the fluorescent light \cite{Chance-1978}. It can differ significantly from unity.  The fraction of the exciton decay rate that is due to fluorescence, or `fluorescence quantum yield' is usually assumed small in organic solar cells. In this paper, we specifically address the case of a large quantum yield and explore the possibilities afforded by this process in improving device performance.

That fluorescence is affected by the device geometry is already well accounted for in the design of organic light-emitting diodes (OLEDs), see for instance \cite{Tsutsui-1991,Saito-1994,Bulovic-1998,Kim-2000,Wasey-2000a,Smith-2005,Flammich-2009,Flammich-2010,Penninck-2012}. For OLEDs, the diffusive transport of excitons is less of an issue than for organic photovoltaics, although this aspect has been considered in \cite{Tessler-2000,Rezzonico-2011}. Indeed, it is obviously desired that OLED excitons fluoresce and they are expected to do so near the interface between donor and acceptor molecules.  For photovoltaic devices, we have recently drawn attention to the usefulness of controlling fluorescence in Schottky solar cells \cite{Vuong-2009}. To this end, we had computed fluorescence rates using Fermi's golden rule, and had therefore neglected dissipative processes, such as non radiative energy transfer to metal electrodes. In this paper, we model excitons as classical dipole emitters,  allowing us to correctly treat dissipation in the electrodes. The radiation of an electromagnetic dipole in a stratified medium has been the subject of many theoretical works \cite{Chance-1978,Lukosz-1980,Michalski-1997,Neyts-1998,Wasey-2000b,Chen-2007,Celebi-2007}, all building from Sommerfeld's analysis of a radio-wave antenna above the Earth \cite{Sommerfeld-1909}. Finally, let us note that the idea of fluorescence management in photovoltaic cells was already put forward in \cite{Becker-1997}, but it was discussed only at a qualitative level.

\begin{figure}
\includegraphics[width=\textwidth]{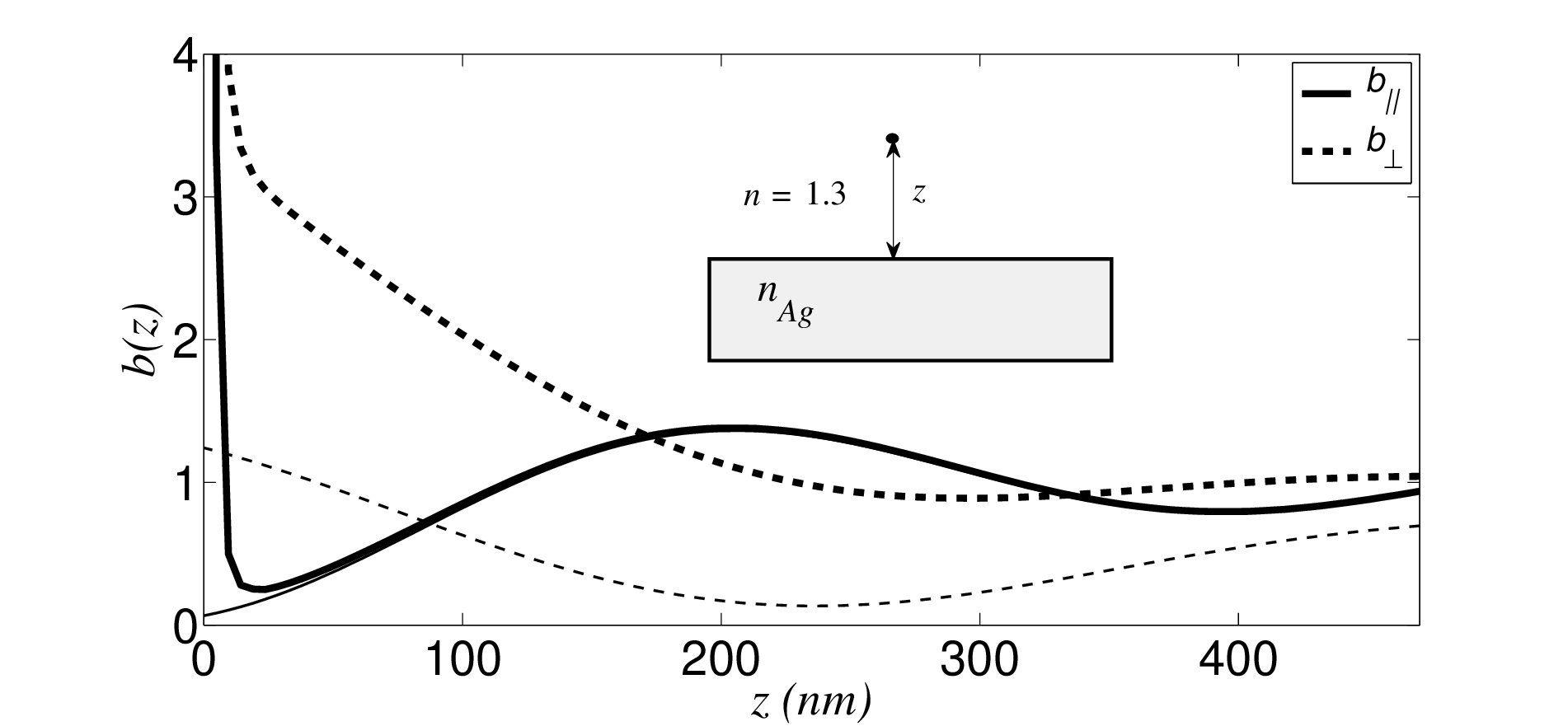}
\caption{Normalized fluorescent decay rate as a function of distance from a thick Ag electrode in a uniform medium of refractive index $n=1.3$. Thin lines: ideal, lossless electrode, $n_\text{Ag}=6.37i$. Thick line: real electrode, $n_\text{Ag}=0.04+6.37i$ \cite{Johnson-1972}. Vaccuum emission wavelength: 900~nm. $b_\|$: parallel excitons. $b_\perp$ perpendicular excitons;}
\label{drexhage}
\end{figure}

One may approach fluorescence management from two limit situations: (i) an emitter inside a perfect, sub-wavelength, cavity and (ii) an emitter in front of a single mirror. From the former, we know \cite{Kleppner-1981} that complete suppression of fluorescence is possible. However, no exterior radiation is admitted inside such a cavity. The latter configuration, being open to the environment, appears to be more relevant to the present discussion. We shall call it the `Drexhage configuration', in reference to Drexhage's famous study of fluorescence near a mirror \cite{Drexhage-1970,Drexhage-1974}. The function $b(z)$ can be obtained from the power radiated by an electromagnetic dipole in this configuration \cite{Sommerfeld-1909,Sommerfeld-1949}. To this end, one should distinguish excitons that are oriented parallel  to the electrode ($b_\|$) from those which are perpendicular to it ($b_\perp$), see Fig.~\ref{drexhage}. 

Let us assume that the  photoactive material has a refractive index $n=1.3$ and that the exciton radiates at a vacuum wavelength of 900~nm. At that wavelength, the refraction index of silver is $n_\text{Ag}=0.04+6.37i$ \cite{Johnson-1972}. If silver were lossless, \text{i.e.}, if $n_\text{Ag}$ were purely imaginary, then it is evident from Fig.~\ref{drexhage} that interesting prospects of device improvements would be afforded by the low values of $b_\|(z)$ near the electrode, $z=0$, and of $b_\perp(z)$ near $z=250$~nm. Note that, since the functions $b_\|(z)$ and $b_\perp(z)$ vary in opposite ways, it appears that taking advantage of space-dependent fluorescence requires one to control exciton orientation. Let us next look at the decay rates in the vicinity of \textit{real} silver. Due to dissipation, the decay rates diverge as the exciton comes close to the electrode. On the one hand, $b_\|(z)$ still has a well-pronounced minimum a few tens of nm away from the electrode. The lower the value of $n$, the lower the minimum, and with $n=1.3$, the minimum of $b_\|(z)\approx 0.25$, representing a doubling of the diffusion length in (\ref{Diff2}). On the other hand, Fig.~\ref{drexhage} shows that perpendicular excitons are affected by dissipation over a significantly longer distance than parallel ones. In this particular configuration, $b_\perp(z)$ is nowhere brought close to zero; micro-cavity effects from more complicated layered geometries are necessary to achieve this.
 
Returning to parallel excitons in Fig.~\ref{drexhage}, we note that the minimum of $b_\|(z)$ is very close to the silver electrode. On the other hand,  $g(z)$, being proportional to sunlight intensity, is also close to zero near the silver electrode due to interference. The advantage of having a small $b(z)$ is thus lost in that situation. Hence, in the search of an optimal design,  both functions $b(z)$ and $g(z)$ must be carefully monitored. This is in line with previous work demonstrating the sensitivity of $g(z)$, and hence device performance, on the geometrical parameters \cite{Pettersson-1999,Betancur-2012}.

At first sight, it may seem that decreasing the spontaneous emission in a solar cell device would also affect stimulated absorption. Indeed, Einstein's coefficients of spontaneous emission and stimulated absorption, respectively $\mathsf{A}$ and $\mathsf{B}$, are related in free space by $\mathsf{A}=(8\pi h \nu^3/c^3) \mathsf{B}$, where $h$ is Planck's constant. In fact, the relation between theses two constants is more generally given by  $\mathsf{A}=M(\nu) h \nu \mathsf{B}$, where $M(\nu)$ is the effective spectral mode density per unit volume. This mode density can be considerably less in a confined environment than in open space, allowing one to reduce spontaneous emission while leaving absorption undisturbed. In the same vein, one should note that, in a solar cell, stimulated absorption is primarily caused by photons coming at normal incidence from the sun, while spontaneous emission occurs in all directions. In the latter process, the details of the resulting electromagnetic field, and hence, the coupling of the emitter to that field, depends on the direction of emission as well as the position within the microcavity. This also affects $M(\nu)$.

The rest of the paper is organised as follows. In section \ref{sec:model}, we discuss the mathematical model (\ref{Diff2}) in more detail. Section \ref{sec:num} presents our numerical results. In \S~\ref{sec:para} we consider a configuration that is favourable to parallel exciton transport, while in \S~\ref{sec:perp} we present a configuration that is suited to perpendicular excitons. Finally, in section \ref{ccl}, we present our conclusions.

\section{Model}\label{sec:model}

\begin{figure}
\centering
\includegraphics[height=8cm]{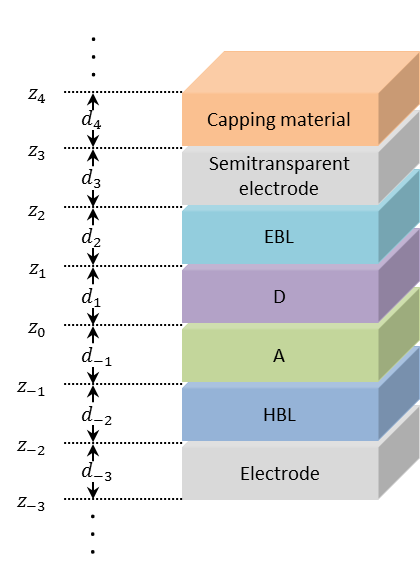}
\caption{Solar cell geometry and coordinate system. HBL: hole-blocking layer. A: Acceptor. D: Donor; EBL: electron-blocking layer.}\label{fig:schema}
\end{figure}

\subsection{Diffusion}
A typical bilayer organic solar cell architecture is sketched in Fig.~\ref{fig:schema}. The photoactive materials occupy the regions $z_{-1}<z<z_0$ and $z_0<z<z_1$. In each of these, the density of excitons, $\rho'$ satisfies \cite{Ghosh-1978}
\begin{align}
0&=D_i\td{^2\rho'}{z^2}-\frac{b(z)}{\tau_i}\rho'+\alpha_i\phi_i N g(z),&
i&=-1,1,
\end{align}
where $D_i$ is the diffusion constant, $\tau_i$ is the bulk value of the exciton lifetime, $\alpha_i$ is the sunlight absorption coefficient, $\phi_i$ the quantum efficiency of exciton generation,  $N$ is the incoming photon flux, and $g(z)$ is the distribution of sunlight intensity inside the device.  $g(z)$ is computed using transfer matrices and assuming a normally incident wave  with unit photon flux at a specific wavelength, $\lambda_s$. Given the complex index of refraction $n_{i,s}$ in region $i$ at $\lambda_s$, the absorption coefficient $\alpha_i$ is given by $4\pi\,\im(n_{i,s})/\lambda_s$.

In each photoactive region, we normalize $\rho'$ as
\beq
\rho' = \alpha_i\phi_iN\tau_i\rho,
\label{adim}
\eeq
giving
\begin{align}
0&=L_i^2\td{^2\rho}{z^2}-b(z)\rho+g(z),&
L_i^2&=D_i\tau_i,
\label{Diff3}
\end{align}
where $\rho$, $b$ and $g$ are dimensionless. We assume that complete exciton dissociation into electrons and holes occurs at the Donor/Acceptor interface. The hole-blocking layer (HBL) is assumed to block exciton current but to let electrons flow easily. Similarly, electron-blocking layers (EBL) are supposed to let holes through but to block exciton current. Hence, (\ref{Diff3}) should be solved subjected to the boundary conditions
\begin{align}
\left.\td\rho z\right|_{z_{-1}}&=0,&
\rho(z_0)&=0,&
\left.\td\rho z\right|_{z_{1}}&=0.
\label{BC}
\end{align}
In the absence of blocking layers, the electromagnetic losses in the electrode make $b(z)$ diverge as $z\to z_{\pm1}$, resulting in a vanishingly small exciton density at these locations, independently of the imposed boundary condition.

Let us note that more detailed transport models than mere exciton diffusion exist, particularly drift-diffusion models \cite{Barker-2003} which include electrons, holes and static electric field distributions within the cell. However, exciton transport asymptotically decouples from the rest of the system and governs the device dynamics in the limit of small exciton mobility \cite{Brinkman-2012}. 

If all excitons reaching the interface at $z_0$ disintegrate into a pair of  electron and hole and all of these reach the electrodes, then  the short-circuit current is the sum of the two diffusion currents at $z_0$. Using (\ref{adim}), it can be expressed as
\begin{align}
I_{\text sc}&=I_{-1}+I_1,&
I_{-1}&= -\alpha_{-1}\phi_{-1}N A \tau_{-1}D_{-1}\left.\td\rho z\right|_{z_0^-},&
I_1&=   \alpha_1\phi_1N A \tau_1D_1\left.\td\rho z\right|_{z_0^+},
\end{align}
$A$ being the area of the interface. We may group the terms in the two diffusion currents above as
\beq
I_i=\pm\left(A N\right)\alpha_i \phi_iL_i^2 \left.\td\rho z\right|_{z_0^{\pm}}.
\eeq
In this expression, $AN$ is the number of photons falling down on the device per unit time, $\alpha_i\phi_i$ is the fraction of them per unit length which are converted into excitons, and the diffusion constants and bulk lifetimes only appear through the diffusion length $L_i$. Let us note that $I_\text{sc}/(AN)$ is the external quantum efficiency (EQE) under the short-circuit condition. We thus have
\beq
\EQE =\alpha_1 \phi_1L_1^2 \left.\td\rho z\right|_{z_0^{+}}-\alpha_{-1} \phi_{-1}L_{-1}^2 \left.\td\rho z\right|_{z_0^{-}}
\eeq
 In this work, we only monitor the short-circuit current as a figure of merit for a given configuration. Indeed, this can be directly obtained from the solution of (\ref{Diff3}), without additional modeling assumptions. 
 
Let us assume that $\phi_i=1$.  We can obtain a useful estimation of $\EQE$ if the photoactive region is sufficiently thin that we can replace $b(z)$ and $g(z)$ by their average values in (\ref{Diff3}):
\begin{align}
0&\approx L_i^2\td{^2\rho}{z^2}-\bar b\rho+\bar g,
&\bar b &=\frac1d\int_{A+D}b(z)\rd z,
&\bar g &=\frac1d\int_{A+D}g(z)\rd z,
\end{align}
where $d$ is the total thickness of the photoactive region. If the Donor and Acceptor material have the same diffusion length $L$, the maximum short-circuit current is obtained when both materials have the same thickness $d/2$. A straightforward calculation then shows that
\beq
\EQE\approx \alpha \bar g d\times \frac{2L}{d\sqrt {\bar b}}\tanh\left(\frac{d\sqrt {\bar b}}{2L}\right)
\equiv\eta_A\times\eta_D
\label{EQE}
\eeq
where $\eta_A$ is the fraction of the photon flux that is absorbed by the photoactive region and $\eta_D$ is a diffusion efficiency, in the same spirit as in \cite{Forrest-2005}.

\subsection{Decay rates}

We now turn to the computation of the normalized decay rate $b(z)$. Its expression was determined for a dipole within an arbitrary stack of parallel layers of isotropic materials by Chance, Prock, and Silbey \cite{Chance-1978}. The case of a dipole embedded in a uniaxial medium sandwiched between two other uniaxial media with aligned extraordinary axes was treated in \cite{Wasey-2000b}. This last case can easily be generalized to an arbitrary stacking of aligned uniaxial media, as we do below. Note that the Green dyadic was computed for a uniaxial multilayered media in \cite{Michalski-1997}, although with no particular emphasis on $b(z)$. For the sake of completeness, we provide an alternative derivation of $b(z)$ to that given in \cite{Wasey-2000b} and give the relevant components of the Green dyadic in appendix to this paper.

When computing the fluorescence in the photoactive layers, it is necessary to assume that the refraction index is purely real in these particular layers. Otherwise, the power required from an electric dipole to sustain harmonic oscillations is infinite and the validity of the classical formulas given below is uncertain. Fortunately, a frequent feature of organic materials is that the exciton energy level is well below the LUMO, with a binding energy ranging between 0.1~eV and 1~eV. Hence, the exciton radiation is only weakly absorbed by the surrounding material and the refraction index at the exciton frequency is mostly real. It appears reasonable, then, to neglect the imaginary part of the refractive index in the donor and acceptor material at the exciton frequency.

Let us assume that the layers in the device are uniaxial and that they all have their extraordinary axis in the $z$ direction. The relative permittivity tensor reads
\begin{align}
\bar\eps_i&=\begin{pmatrix}
\eps_{i,x}&0&0\\ 0&\eps_{i,x} &0\\0&0&\eps_{i,z}
\end{pmatrix},&
z_{i-1}<z<z_i.
\label{permittivity}
\end{align}
Given the vacuum wavenumber $k_0$, the dispersion relation for extraordinary waves, or $p$-waves, is
\beq
k_{e,z,i}=\sqrt{\eps_{i,x}k_0^2-\frac{\eps_{i,x}}{\eps_{i,z}}k_\|^2},
\label{extra}
\eeq
between the $z$-component,  $k_{e,z,i}$, of the  wave vector and its projection $k_\|$  on the ($x,y$) plane. On the other hand, ordinary waves, or $s$-waves, satisfy the dispersion relation
\beq
k_{o,z,i}=\sqrt{\eps_{i,x}k_0^2-k_\|^2}.
\label{ordi}
\eeq
With these notations in mind, the normalized decay rate for an exciton oriented in the $z$ direction, and located between $z_0$ and $z_1$, is given by (see \cite{Wasey-2000b} or the Appendix)
\begin{align}
b_\perp(z)=
1+\frac{3q}2\frac{\eps_{1,x}^{1/2}}{\eps_{1,z}^2}\re\left\{\frac1{k_0^3}\int_0^\infty
\frac{\hat R_0^p+\hat R_1^p+2\hat R_0^p \hat R_1^p}{1-\hat R_0^p\hat R_1^p}\frac{k_\|^3}{k_{e,z,1}}\rd k_\|\right\},
\label{bperp}
\end{align}
where $q$ is the fluorescence quantum yield,
\begin{align}
\hat R_0^p&=R_0^p \exp\left[2ik_{e,z,1}\left(z-z_0\right)\right],
&\hat R_1^p&=R_1^p \exp\left[2ik_{e,z,1}\left(z_1-z\right)\right],
\end{align}
and $R_0^p$ and $R_1^p$ are the coefficients of reflection from the layers below and above the exciton, respectively. 

To compute the reflection coefficient $R_i^p$ of a $p$-wave incident from a medium with permittivity tensors $\bar \eps_{i}$ onto a multilayered medium with permittivity tensors $\bar \eps_{i+1}$, $\bar\eps_{i+2},\ldots, \bar\eps_{N}, \bar\eps_{N+1}$ and thicknesses $d_{i+1}$, $d_{i+2},\ldots,d_{N}$ we use the downwards recurrence
\begin{align}
R_{N}^p&=R_{N,N+1}^p,\\
R_{j-1}^p&=\frac{R_{j-1,j}^p+  R_{j}^p \exp\left(2ik_{e,z,j}d_j\right)}{1+R_{j-1,j}^p  \;R_{j}^p \exp\left(2ik_{e,z,j}d_j\right)},
\label{recurrence}
\end{align}
where  $R_{ij}^p$ is the $p$-wave reflection coefficients between half-spaces filled with media $i$ and $j$:
\beq
R_{ij}^p=\frac{k_{e,z,i}\eps_{j,x}-k_{e,z,j}\eps_{i,x}}{k_{e,z,i}\eps_{j,x}+k_{e,z,j}\eps_{i,x}}.
\eeq

For an exciton with dipole moment in the $(x,y)$ plane, and $z$ between $z_0$ and $z_1$, the decay rate is given by the formula
\begin{multline}
b_\|(z)=1+\frac{3q}{4}\frac{4\eps_{1,x}^{1/2}}{3\eps_{1,x}+\eps_{1,z}}\\\times
\re\left\{\frac1{k_0}
\int_0^\infty\left(
\frac{k_{e,z,1} }{\eps_{1,x} k_0^2}\frac{2\hat R_0^p\hat R_1^p-\hat R_0^p-\hat R_1^p}{1-\hat R_0^p\hat R_1^p}
+\frac{1}{k_{o,z,1}} \frac{2\hat R_0^s\hat R_1^s+\hat R_0^s+\hat R_1^s}{1-\hat R_0^s\hat R_1^s}
\right)k_\|\rd k_\|
\right\}.
\label{bpara}
\end{multline}
Here, 
\begin{align}
\hat R_0^s&=R_0^s \exp\left[2ik_{o,z,1}\left(z-z_0\right)\right],
&\hat R_1^s&=R_1^s \exp\left[2ik_{o,z,1}\left(z_1-z\right)\right],
\end{align}
where $R_0^s$ and $R_1^s$ are computed by a similar recurrence as for $p$-waves:
\begin{align}
R_{N}^s&=R_{N,N+1}^s,&
R_{j-1}^s&=\frac{R_{j-1,j}^s+ R_{j}^s \exp\left(2ik_{o,z,j}d_j\right)}{1+R_{j-1,j}^s\; R_{j}^s \exp\left(2ik_{o,z,j}d_j\right)},&
R_{ij}^s=\frac{k_{o,z,i}-k_{o,z,j}}{k_{o,z,i}+k_{o,z,j}}.
\end{align}
The computation of $b_\perp$ and $b_\|$ in the region $z_{-1}<z<z_0$ follows the same pattern. 

Finally, for randomly oriented excitons, $b=\frac23b_\|+\frac13b_\perp$.


\section{Numerical results}\label{sec:num}

We solved Eq.~(\ref{Diff3}) in the two active regions depicted in Fig.~\ref{fig:schema} for various device configurations.  The bottom electrode was assumed to be Ag in all cases and the transparent electrode was taken to be  either ITO or a thin Ag layer. Note that other metals, such as Au or Al, could equally be used in the simulations; from an electromagnetic point of view, they influence the device performance in the same way as Ag. The refraction indexes of Ag and ITO were taken from \cite{Johnson-1972} and  \cite{refractiveindex.info}, respectively. The rate of production of excitons, $g(z)$, was computed at a wavelength $\lambda_s=750$~nm, where the solar photon flux is highest. On the other hand, fluorescence was computed at a wavelength $\lambda=900$~nm, corresponding to an exciton binding energy of 0.275~eV. For each simulation, the layer thicknesses must be specified as well as their refractive indexes at $\lambda$ and $\lambda_s$. For the exciton- and hole-blocking layers, we assumed identical real refraction indexes at $\lambda$ and $\lambda_s$, ranging between 1.3 and 2.8. In the Donor and Acceptor materials, an imaginary part was added to the refraction indexes at $\lambda_s$ in order to account for a prescribed sunlight absorption coefficient. In all cases, we assumed an absorption length, $\alpha^{-1}$, of 70 nm. In addition, the quantum yield and the diffusion length should be prescribed in each photo-active layer. Having so many parameters to vary, we imposed $\eps_x=\eps_z$ everywhere despite the anisotropy implied by assuming either $b=b_\|$ or $b=b_\perp$. Indeed, numerical evaluations of the function $b(z)$ at the end of \cite{Chance-1978} and in \cite{Wasey-2000a} suggests to us that a moderate anisotropy would only slightly modify an optimal geometry found in the isotropic limit. We leave, therefore, the full investigation of anisotropic effects for future research. Finally, the quantum efficiency of exciton generation is taken to be unity in both donor and acceptor regions: $\phi_i=1$.

Given geometrical parameters and refraction indexes at $\lambda_s$, the function $g(z)$ can easily be computed with standard transfer matrix methods \cite{BornWolf,Pettersson-1999}. Assuming a unit-amplitude wave $\exp\left( -in_\text{ext}k_0z\right)$ coming from $z=\infty$, one finds the wave $a_i \exp\left(-i n_{i,s}k_0z\right)+b_i \exp\left(i n_{i,s}k_0z\right)$ in region $i$ and 
\beq
g(z)=\frac{\re \left(n_{i,s}\right)}{n_\text{ext}}\left|a_i e^{-i n_{i,s}k_0z}+b_i e^{i n_{i,s}k_0z}\right|^2.
\label{eqg}
\eeq
The numerical evaluation of integrals of the type which appears in  (\ref{bperp}) and (\ref{bpara}) may present some difficulties when the denominators of the integrand approach zero. To avoid this problem, it was recommended to deform the integration contour away from these poles \cite{Chen-2007}. This is especially interesting when a large number of evaluations are required and when the poles are numerous. In our case, we checked that it was unnecessary, as the number of poles was small and that they were in practice sufficiently far from the real axis. More troublesome, numerically, can be the square root singularities at $k_{e,z,1}=0$ and $k_{o,z,1}=0$ (both expressions being identical if $\eps_x=\eps_z$.) These are simply removed by changing the integration variable from $k_\|$ to either $k_{e,z,1}$ or $k_{o,z,1}$. Furthermore, it is useful to adopt the wavenumber unit such that $k_0=1$. Regarding the integration of (\ref{Diff3}), we discretized each photoactive region with a Chebychev grid \cite{Trefethen-2000}. In this way $\EQE$ was accurately computed with significantly fewer discretization points, and hence fewer evaluations of $b(z)$, than with standard finite differences.

\subsection{Parallel excitons}\label{sec:para}

\begin{figure}
\centering
\includegraphics[width=1\textwidth]{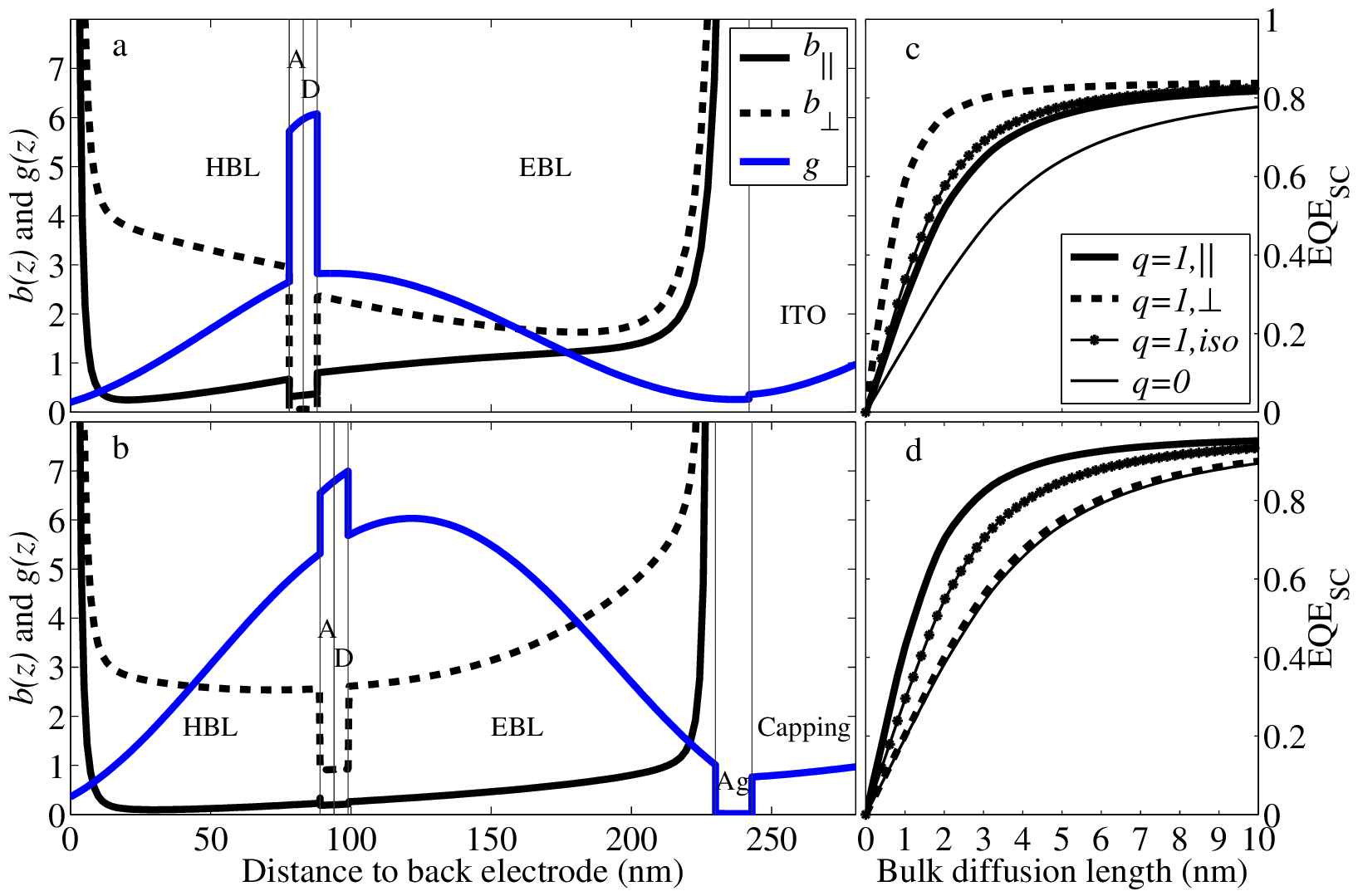}
\caption{Decay rates ($q=1$), gain functions, and external quantum efficiencies for the parameters given in Table~\ref{tablepara}.  Top: ITO transparent electrode. Bottom: Ag transparent electrode. iso: isotropic case where $b=\frac23b_\|+\frac13b_\perp$.}
\label{paracase}
\end{figure}

\begin{table}\centering\small
\caption{\label{tablepara}{\bf Two configurations optimized for parallel excitons. A sunlight absorption length of 70 nm is assumed in both photoactive materials.} $n_\text{ITO}=1.76+0.08i$, $n_\text{Ag}(900\text{nm})=0.04+6.37i$, $n_\text{Ag}(750\text{nm})=0.03+5.19i$. }
\begin{tabular}{cccc|cccc}
\hline
layer & thickness  & $n(\lambda)$  & $n(\lambda_s)$  
&layer & thickness & $n(\lambda)$ & $n(\lambda_s)$ \\
&(nm)&&&&(nm)&&
\\
\hline
air &$\infty$ &1&1 &air &$\infty$ &1&1\\
capping & 56 &1.3 &1.3				    &capping & 123 &1.3 &1.3\\
ITO    & 74 & $n_\text{ITO}$ & $n_\text{ITO}$   &Ag      & 13       &$n_\text{Ag}$ & $n_\text{Ag}$ \\
EBL    & 154 	& 1.3 & 1.3                                     &EBL    & 131 	& 1.3 & 1.3 \\
D       & 5 	& 2.8 & $2.8 + 0.85i$		    &D       & 5 	& 1.6 & $1.6 + 0.85i$ \\
A       & 5 	& 2.8 & $2.8 + 0.85i$ 		    &A       & 5 	& 1.6 & $1.6 + 0.85i$ \\
HBL    & 78 	& 1.3 & 1.3 				    &HBL    & 89 	& 1.3 & 1.3 \\
Ag    & $\infty$& $n_\text{Ag}$ & $n_\text{Ag}$&Ag    & $\infty$  	& $n_\text{Ag}$ & $n_\text{Ag}$ \\
\hline
\end{tabular}
\end{table}

Based on the geometry considered in the introduction, we first look for a structure that is favorable to parallel excitons. It was noted that a low refractive index next to  silver can bring $b_\|(z)$ to a minimum near the back electrode, but that $b_\|(z)$ rapidly increases further away from it. It is therefore desirable to extend the range of low $b_\|$ values away from the back electrode so as to combine long lifetime with good sunlight harvesting. This is achieved with the configurations described in Table~\ref{tablepara}. One configuration uses ITO as the transparent electrode, the other one uses Ag. The parameters were determined to maximize $\EQE$ for $q=1$ and a bulk diffusion length of 5~nm. Figs.~\ref{paracase}a and b show the corresponding functions $b_\|(z)$, $b_\perp(z)$, and $g(z)$. One may see that, indeed, low $b_\|$ values are maintained over a longer distance from the back electrode than in Drexhage's configuration. Moreover, and contrary to intuition, Fig.~\ref{paracase}a gives an example where both $b_\|$ and $b_\perp$ are small at the same place. Solving (\ref{Diff3}) and (\ref{BC}) with these parameters and various bulk diffusion lengths, $L$, we plot the $\EQE$ as a function of $L$ in Figs.~\ref{paracase}c and d. This allows one to compare the merit of non-fluorescent ($q=0$) and purely fluorescent ($q=1$) materials with parallel, perpendicular or randomly oriented excitons. The theoretical advantage of $q=1$ appears clearly, even in the isotropic case. This is particularly true when the diffusion length is small compared to the thickness of the active region. Most of the $q=1$ curves have almost reached their asymptotic value for $L=5$ nm while twice that diffusion length is necessary for $q=0$. Hence, the effective diffusion length is approximately doubled in these examples thanks to fluorescence.

\begin{figure}\centering
\includegraphics[width=.8\textwidth]{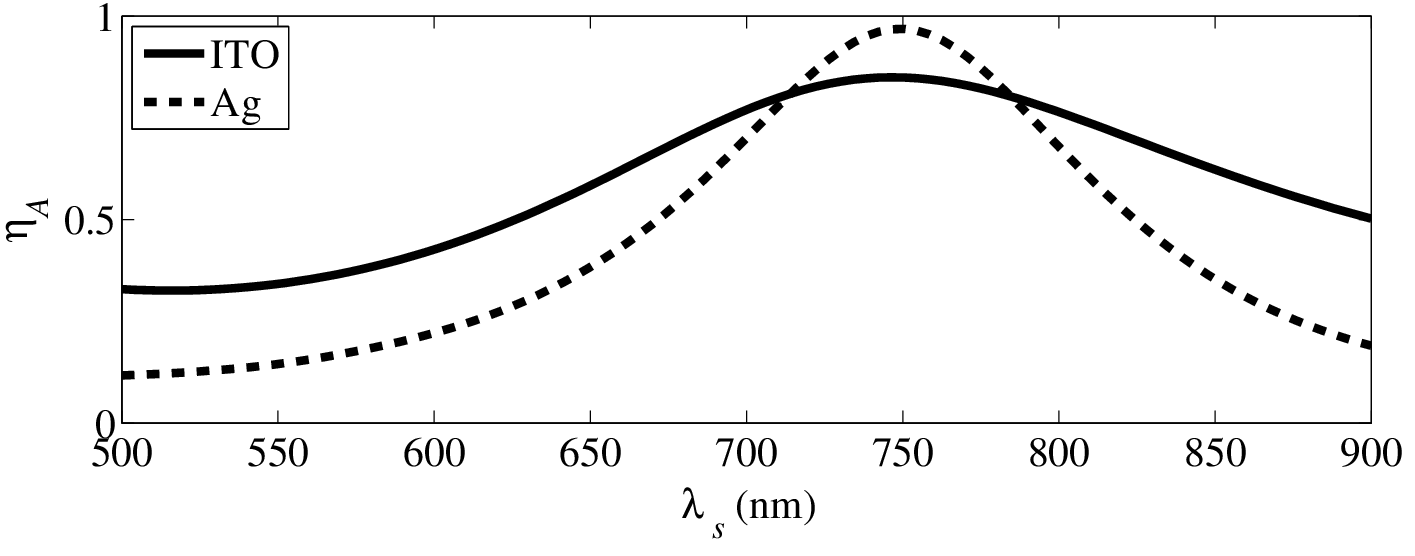}
\caption{Absorption efficiency $\eta_A$ computed for the devices described in Table~\ref{tablepara} as a function of sun wavelength. A fixed ITO refractive index $n_\text{ITO}=1.76+0.08i$ and an absorption length of 70 nm in the active region are assumed over the whole spectral range. Ag refractive index is taken from \cite{Johnson-1972}.}
\label{Eta_AvsLambda}
\end{figure}

All the curves $\EQE(L)$ displayed are very well approximated by (\ref{EQE}) with suitable values of $\bar b$ estimated from Figs~\ref{paracase}a,b. The ultimate $\EQE$, as $L$ becomes large, is given by the absorption efficiency, $\eta_A$. At the wavelength considered, the use of an Ag transparent electrode appears advantageous compared to ITO, as it allows one to obtain large values of $\eta_A$. If one varies $\lambda_s$ over some tens of nm, however, one witnesses a sharper drop of $\eta_A$ with Ag than with ITO. Hence, a silver transparent electrode makes light injection more resonant than an ITO electrode. This is illustrated in Fig.~\ref{Eta_AvsLambda}.

\FloatBarrier
\subsection{Perpendicular excitons}\label{sec:perp}

\begin{table}[h!]\centering\small
 \caption{\label{tableperp}{\bf Optimized configuration for perpendicular excitons. Absorption length: 70 nm is assumed in both photoactive materials.} $n_\text{ITO}=1.76+0.08i$, $n_\text{Ag}(900\text{nm})=0.04+6.37i$, $n_\text{Ag}(750\text{nm})=0.03+5.19i$. }
\begin{tabular}{cccc}
\hline
layer & thickness  & $n(\lambda)$ & $n(\lambda_s)$ \\
&(nm)&&
\\
\hline
air &$\infty$ &1&1\\
capping & 94 &1.8 &1.8\\
Ag     & 10     &  $n_\text{Ag}$ & $n_\text{Ag}$ \\
EBL    & 127 	& 1.3 & 1.3 \\
D       & 5 	& 2.5 & $2.5 + 0.853i$ \\
A       & 5 	& 2.5 & $2.5 + 0.853i$ \\
HBL    & 89 	& 1.3 & 1.3 \\
Ag    & $\infty$  	& $n_\text{Ag}$ & $n_\text{Ag}$ \\
\hline
\end{tabular}
\end{table}
\begin{figure}\centering
\includegraphics[width=\textwidth]{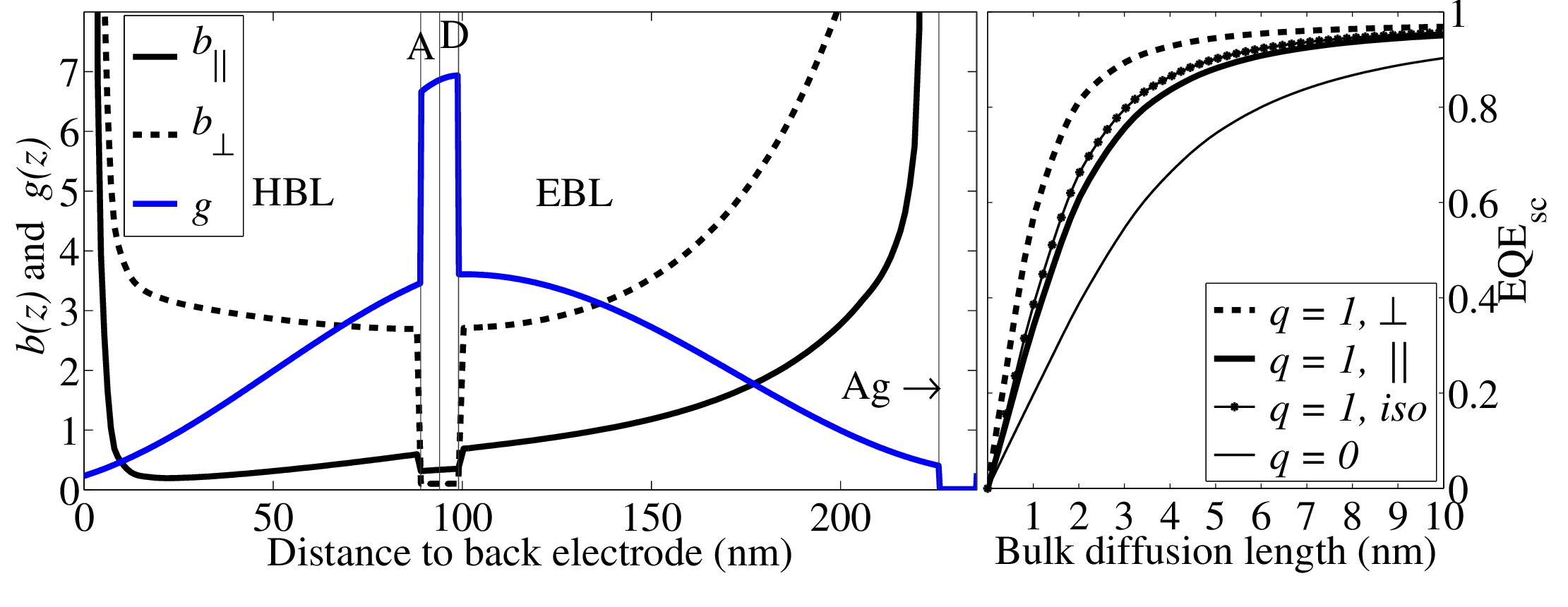}
\caption{Gain and decay profiles computed with Table~\ref{tableperp} and $q=1$. HBL: hole-blocking layer, A: acceptor, D: donor, EBL: electron-blocking layer.}
\label{perpcase}
\end{figure}

We now turn to the transport of perpendicular excitons. Considering the formula for $b_\perp(z)$ in~(\ref{bperp}), we note that large absolute values of $R_0^p$ and $R_1^p$ are required in order to significantly alter the decay rate. A way to favor such a situation is to raise the refraction index in the photoactive region. On the other hand, a large index contrast may induce a strong reflection of the sunlight at the D/EBL interface, which is detrimental to the gain function $g(z)$.  A compromise between the two effects is given in Table~\ref{tableperp}. As with parallel excitons, we find that a metallic transparent electrode can give rise to a higher short-circuit current than with ITO and we thus focus on a silver electrode. The gain and loss profiles corresponding to Table~\ref{tableperp} are shown in Fig.~\ref{perpcase}. Remarkably  small decay rates are obtained in the photoactive region, resulting in a dramatic improvement of $\EQE$ when passing from $q=0$ to $q=1$. 

\begin{figure}\centering
\includegraphics[width=.8\textwidth]{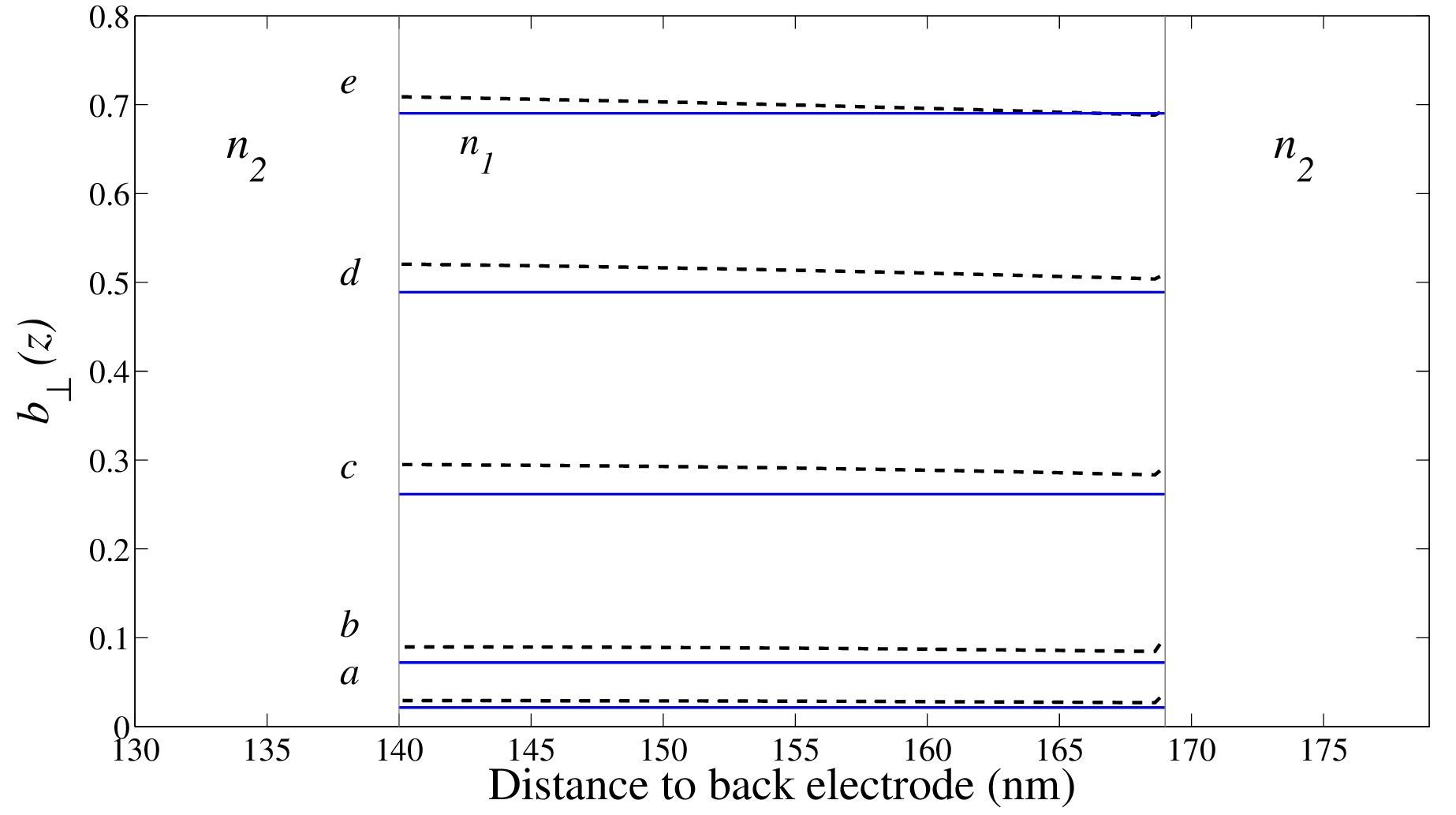}
\caption{$b_\perp(z)$ in a structure with refractive indexes $n_\text{Ag}/n_2/n_1/n_2/n_\text{Ag}/n_3$ and thicknesses~(nm): $140/ 29/ 295 / 6$. $n_2=1.3$, $n_3=2.8$. Cases $a$ to $e$: $n_1=2.8, 2.2, 1.7, 1.5, 1.4$. Wavelength: 900 nm. Dashed line: numerical. Full line: $(n_2/n_1)^5$.
}
\label{fifthpower}
\end{figure}

The very small decay rates shown in Fig.~\ref{perpcase} and in Fig.~\ref{paracase}a motivate us to try and find a simple estimate of the minimum achievable $b_\perp$  in a solar cell. To this end, let us assume a high-index photoactive material ($n_1$) between low-index blocking layers ($n_2$). Restricting our attention to isotropic media, we have $\eps_{i,x}=\eps_{i,z}=n_i^2$ and $k_{e,z,i}=k_{o,z,i}\equiv k_{z,i}$. In particular,
\begin{align}
k_{z,1}&=\sqrt{n_1^2k_0^2-k_\|^2}, 
& k_{z,2}&=\sqrt{n_2^2k_0^2-k_\|^2}.
\label{iso}
\end{align}
On account of the thinness of the photoactive region, we neglect the phase factors in  $\hat R_0^p$ and  $\hat R_1^p$ and approximate them by $R_0^p$ and $R_1^p$, respectively. Furthermore, we assume that the electrodes are sufficiently remote for their influence to be negligible. This point is partly justified for values of $k_\|$ in (\ref{bperp}) such that $k_{z,2}$ is imaginary but is not otherwise rigorous. Thus, we write
\beq
\hat R_0^p\approx\hat R_1^p\approx R=\frac{k_{z,1}n_2^2-k_{z,2}n_1^2}{k_{z,1}n_2^2+k_{z,2}n_1^2},
\eeq
and the expression for $b_\perp$ becomes
\begin{align}
b_\perp(z)&\approx
1+\frac{3q}{2n_1^3} \re\left\{\frac1{k_0^3}\int_0^\infty \frac{ 2   R}{1-R}\frac{k_\|^3}{k_{z,1}}\rd k_\|\right\},\\
&=
1+\frac{3q}{2n_1^3} \re\left\{\frac1{k_0^3}\int_0^\infty 
\frac{k_{z,1}n_2^2-k_{z,2}n_1^2}{k_{z,1}k_{z,2}n_1^2}
 k_\|^3 \rd k_\|\right\},\\
 &=
1+\frac{3q}{2n_1^3} \re\left\{\frac1{k_0^3}\int_0^\infty 
\left(\frac{n_2^2/n_1^2}{k_{z,2}}-\frac{1}{k_{z,1}}\right)
 k_\|^3 \rd k_\|\right\}
\end{align}
Using (\ref{iso}) and noting that $x^3/\left(a-x^2\right)^{1/2}=-\frac{1}{3}\td{}{x} \left[\left(a-x^2\right)^{1/2}\left(2a+x^2\right)\right]$, the above integral can be evaluated analytically, yielding
\beq
b_\perp\approx1-q+q\left(\frac{n_2}{n_1}\right)^5.
\label{approx}
\eeq
In particular, for unit fluorescence quantum yield, we have the approximate formula $b_\perp\approx\left(n_2/n_1\right)^5$. As shown in Fig.~\ref{fifthpower}, this simple expression provides a fairly good approximation for the whole range of index contrast allowed by $1.3<n_2<n_1<2.8$. With $n_1=2.8$ and $n_2=1.3$, the minimal achievable decay rate is estimated as $b_{\perp,\min}~\approx~0.022$, to be compared with the actual value, 0.027, in Fig.~\ref{fifthpower}. Moreover, (\ref{approx}) remains a good estimate as long as the distance to the nearest silver electrode exceeds 100~nm. Below that distance, energy transfer to the silver electrodes, notably through surface plasmon polaritons \cite{Amos-1997}, becomes important. A remarkable feature of the approximation (\ref{approx}) is that it is independent of the exciton wavelength and of geometrical parameters. 
 
\section{Conclusion and perspectives}\label{ccl}

Next to the usual material characteristics such as absorption spectrum, LUMO and HOMO, exciton binding energy, lifetimes and mobilities, this study suggests to also consider fluorescence quantum yield, $q$, in the quest for good photovoltaic molecules. Molecules with a high quantum yield are amenable to fluorescence management, by which the exciton diffusion length can be considerably increased. The approximate formula (\ref{EQE}) indicates that, for thin photoactive regions,
\beq
\EQE =\eta_A(\lambda_s)\eta_D(\lambda, q),
\eeq
which brings out the necessity to jointly consider the light harvesting capability ($\eta_A$) and the transport property of a solar cell ($\eta_D$). Both depend on the geometry, but fluorescence affects $\eta_D$ only.

Our study contradicts several expectations derived from Drexhage's configuration. To start with, the  strongest fluorescence inhibition was found for perpendicular excitons, rather than for parallel ones. An approximate rule is that perpendicular exciton lifetime can be increased up to a factor $(n_1/n_2)^5$, where $n_1$ is the refraction index in the photoactive material and $n_2$ is the index in the neighboring layers. This amounts to increasing the diffusion length by a factor $(n_1/n_2)^{5/2}$ with respect to the bulk value. This formula holds well if the HBL and EBL are sufficient thick for dissipative energy transfer to be negligible and if the active region is thin compared to the fluorescence wavelength. Secondly, by an appropriate choice of the optical and geometrical parameters, one can have reduced fluorescent decay rate for perpendicular and parallel excitons at the same places. Hence, improvement of device performance is theoretically possible for $q=1$ also with random exciton orientation. 

The ultimate $\EQE$ is $\eta_A$. In this respect, a metallic transparent electrode such as Ag was found to yield larger values than a dielectric one (at optical frequencies), such as ITO. Moreover, in the example given in Table~\ref{tablepara}, an Ag thickness of 13~nm was found to be preferable than the more usual 10~nm. Indeed, the resulting fluorescence inhibition and improved absorption in the photoactive region at $\lambda_s=750$ nm more than compensate absorption losses in the transparent electrode incurred by passing from 10 to 13 nm thickness.  At the same time $\eta_A(\lambda_s)$ shows a sharper spectral resonance with Ag than with ITO, see Fig.~\ref{Eta_AvsLambda}. For simplicity, Fig.~\ref{Eta_AvsLambda} was drawn by assuming a fixed absorption length over the whole frequency range considered. In practice, the most judicious choice between ITO and a metallic transparent electrode should involve matching the microcavity response with the molecular absorption spectrum.

The optimal geometries presented in this paper are slightly unusual, particularly regarding the large thicknesses of the HBL and EBL.  Such thicknesses are not uncommon in OLED design. Presently, they are necessary to make fluorescent excitons immune to dissipative energy transfer in the electrodes. 

Our work also shows that, for a badly designed geometry, a purely fluorescent material is less efficient than a non-fluorescent material with the same bulk lifetime. This can easily be seen by considering Fig.~\ref{perpcase}. If the gain material was located between 150 and 200 nm in that example, then both $b_\perp$ and $b_\|$ would be larger than one. In that case, obviously, $\eta_D(\lambda,q=1)< \eta_D(\lambda,q=0)$. This may be surprising, as  Shockley and Queisser showed that a large quantum yield was necessary to maximize the solar cell efficiency \cite{Shockley-1961,Miller-2012}. However, the Shockley-Queisser limit was derived for the much thicker semiconductor devices, where the microcavity effects discussed here are negligible and where exciton transport is not an issue.

In this work, we have only made a partial exploration of the parameter space. There is therefore still room for improvement or adaptation to technical realities. In the frame of the present theory, the number of configurations that can be studied exceeds by far the examples that we have discussed. We have limited our parameter investigation to isotropic materials, to donor and acceptor molecules having unit quantum yield, identical diffusion lengths, and identical orientation. These constraints can obviously be lifted with the mathematical model presented here. Besides, other sunlight injection strategies could possibly be devised that combine large value of $\eta_A$ with large $\eta_D$. 

Finally, although we focused on large-$q$ molecules, the present theory should also be useful to model materials with moderate values of $q$. Indeed, the functions $b_\perp$ and $b_\|$ can still vary significantly from unity in that case. Models like (\ref{Diff2}) then appear to be more sensible than (\ref{Diff1}), while still simple to implement. In particular, important experimental estimations of diffusion lengths based on the assumption of  spatially uniform exciton lifetimes may have to be revised for some molecules.

\section*{Acknowledgments}
G.K. is a Research Associate of the Fonds de la Recherche Scientifique - FNRS (Belgium.)

\appendix
\section{Appendix: Derivation of  (\ref{bperp}) and (\ref{bpara})}

In this appendix, we derive formula (\ref{bperp}) and (\ref{bpara}) for a uniaxial multilayer. As a preample, let us note that if the purely fluorescent part of the exciton decay rate can be written as
\beq
\Gamma_r\left(1+\gamma(z)\right),
\eeq
where $\Gamma_r$ is the bulk value and $\gamma(z)$ is the correction due to the boundaries, then the total decay rate is
\beq
\Gamma_{nr}+\Gamma_r\left(1+\gamma(z)\right)=\left(\Gamma_{nr}+\Gamma_r\right)\left(1+\frac{\Gamma_r}{\Gamma_{nr}+\Gamma_r}\gamma(z)\right).
\eeq
Therefore,
\beq
b(z)=1+q\gamma(z).
\eeq
Considering time-harmonic electromagnetic fields, Amp\`ere's and Faraday's law can respectively be written as
\begin{align}
k_0^2 \bar\eps\ve E&=-i\omega\mu\ve J+i\omega\curl\ve B,\label{curlB}\\
i\omega\ve B&=\curl\ve E,\label{curlE}
\end{align}
where $k_0=\sqrt{\eps_0\mu}\omega$ and $\bar\eps$ is the permittivity tensor, of the form given in (\ref{permittivity}). Note that (\ref{curlE}) automatically makes $\ve B$ divergence-free and that $\Div\ve D=\rho$ follows from (\ref{curlB}) and the continuity equation $\Div\ve J-i\omega\rho=0$. We find it useful, as in \cite{Sullivan-1997,Jackson-1999}, to use the $z$ components of $\ve E$ and $\ve B$ as electromagnetic potentials. We shall decompose any vector $\ve Z$ into its $z$ component and its ($x,y$) projection as
\beq
\ve Z=\ve Z_\|+\ez \,Z_z.
\eeq
After some manipulations, we find that the plane components of (\ref{curlB}) and (\ref{curlE}) can be written as
\begin{align}
\left(\pd{^2}{z^2}+ \eps_xk_0^2\right)\ve E_\|&=-i\omega\mu\ve J_\|+\grad_\|\pd{E_z}z-i\omega\ez\times\grad_\|B_z,\label{curlBt}\\
\left(\pd{^2}{z^2}+\eps_x k_0^2 \right)\left(i\omega\ve B_\|\right)&=-i\omega\mu\pd{}z\left(\ez\times\ve J_\|\right)+i\omega\grad_\|\pd{B_z}z-k_0^2\eps_x\ez\times\grad_\| E_z,\label{curlEt}
\end{align}
while
\begin{align}
\left(\eps_z\pd{^2}{z^2}+\eps_x\nabla^2_\|+\eps_x\eps_zk_0^2\right)E_z&=-i\omega\mu\eps_x\ez\cdot\left( \ve J+\frac1{\eps_xk_0^2}\grad\Div\ve J \right),\label{Ez}\\
\left(\pd{^2}{z^2}+\nabla^2_\|+\eps_xk_0^2\right) B_z&=-\mu\ez\cdot\left(\curl\ve J\right).\label{Bz}
\end{align}
Equations (\ref{Ez}) and (\ref{Bz}) make it apparent that $E_z$ is the electromagnetic potential for extraordinary waves and that $B_z$ is the potential for ordinary waves.
Note that the left-hand-side of (\ref{Ez}) can be transformed into the Helmholtz equation by rescaling the coordinates. Therefore,  the Green's functions for Eqs.~(\ref{Ez}) and (\ref{Bz}) in free space are, respectively \cite{Jackson-1999},
\begin{align}
g_e&=\frac{-e^{\,ik_0\sqrt{\eps_x(x^2+y^2)+\eps_zz^2}}}{4\pi\eps_x^{1/2}\sqrt{\eps_x(x^2+y^2)+\eps_zz^2}},&
g_o&=\frac{-e^{\,i\eps_x^{1/2}k_0\sqrt{ x^2+y^2+ z^2}}}{4\pi\sqrt{x^2+y^2+ z^2}}.
\end{align}
These functions have the following representation, due to Sommerfeld \cite{Sommerfeld-1949}:
\begin{align}
g_e&=\frac{-i}{4\pi\eps_z}\int_0^\infty\frac{J_0(k_\|\rho)}{k_{e,z}}e^{ik_{e,z}|z|}k_\|\rd k_\|,&
g_o&=\frac{-i}{4\pi}\int_0^\infty\frac{ J_0(k_\|\rho)}{k_{o,z}}e^{ik_{o,z}|z|}k_\|\rd k_\|,
\label{Somm}
\end{align}
where $\rho=\sqrt{x^2+y^2}$ and $k_{e,z}$, $k_{o,z}$ are given by the dispersion relations (\ref{extra}) and (\ref{ordi}) in the medium being considered. 

With  $\ve J=\ve j_0 \delta(\ve x)$, we may now write the solution of (\ref{Ez}) and (\ref{Bz}) as
\begin{align}
E_z&=-i\omega\mu\eps_x\left(\ez\cdot\ve j_0+\frac1{\eps_xk_0^2} \pd{}z\,\ve j_0\cdot\grad\right)g_e+E',&
B_z&=i\mu\ez\cdot\left(\ve j_0\times\grad\right)g_o+B',
\label{general}
\end{align}
where $E'$ and $B'$ are non singular solutions of the homogenous differential problem that ensure that  boundary conditions at finite distance are satisfied.

\subsection{Perpendicular dipole}
Let us assume first that $\ve j_0=j_0\ez$. We may then take $B_z=0$ and, combining (\ref{Somm}) and (\ref{general}), write $E_z$ as
\beq
E_z=\frac{-\omega\mu\eps_x j_0}{4\pi\eps_z^2k_0^2}\int_0^\infty \frac{k_\|^3}{k_{e,z}}J_0(k_\|\rho)
\left(e^{ik_{e,z}|z|}+Ce^{ik_{e,z}z}+De^{-ik_{e,z}z}\right)\rd k_\|,
\label{Ezperp}
\eeq
where $C$ and $D$ are arbitrary functions of $k_\|$. Let the nearest interfaces be at a distance $\ell_1$ above and a distance $\ell_0$ below the dipole; these are $z_1-z$ and $z-z_0$ in (\ref{bperp}) and (\ref{bpara}). The boundary conditions at these interfaces translate into reflection coefficients for the waves that compose the integral in (\ref{Ezperp}). 
For a given $k_\|$, a wave with amplitude  $(1+C)$ travels upwards from the origin. It is reflected with a coefficient $R_1^p$ at $z=\ell_1$, to be computed according to (\ref{recurrence}). 
Hence, $D$ should be
\beq
D=R_1^pe^{2ik_{e,z}\ell_1}(1+C)\equiv \hat R_1^p(1+C).
\label{eqD}
\eeq
By the same token, considering waves traveling downwards from the origin to the interface at $z=-\ell_0$, the amplitude $C$ is given by
\beq
C=R_0^pe^{2ik_{e,z}\ell_0}(1+D)\equiv \hat R_0^p(1+D).
\label{eqC}
\eeq
Solving (\ref{eqD}) and (\ref{eqC}), we thus find that 
\begin{align}
C&=\frac{\hat R_0^p\hat R_1^p+\hat R_0^p}{1-\hat R_0^p\hat R_1^p},&
D&=\frac{\hat R_0^p\hat R_1^p+\hat R_1^p}{1-\hat R_0^p\hat R_1^p},
\end{align}
and, hence, that
\beq
E_z=\frac{-\omega\mu \eps_xj_0}{4\pi\eps_z^2k_0^2}\int_0^\infty \frac{k_\|^3}{k_{e,z}}J_0(k_\|\rho)
\left(e^{ik_{e,z}|z|}+\frac{\hat R_0^p\hat R_1^p+\hat R_0^p}{1-\hat R_0^p\hat R_1^p}e^{ik_{e,z}z}+
\frac{\hat R_0^p\hat R_1^p+\hat R_1^p}{1-\hat R_0^p\hat R_1^p}e^{-ik_{e,z}z}\right)\rd k_\|.
\eeq
With this solution, Eq.~(\ref{curlBt}) for $\ve E_\|$ becomes
\beq
\left(\pd{^2}{z^2}+ \eps_xk_0^2\right)\ve E_\|=
\frac{-\omega\mu\eps_x j_0}{4\pi\eps_z^2k_0^2}\grad_\|\pd{}z\int_0^\infty \frac{k_\|^3J_0(k_\|\rho)}{k_{e,z}} 
\left(e^{ik_{e,z}|z|}+\frac{\hat R_0^p\hat R_1^p+\hat R_0^p}{1-\hat R_0^p\hat R_1^p}e^{ik_{e,z}z}+
\frac{\hat R_0^p\hat R_1^p+\hat R_1^p}{1-\hat R_0^p\hat R_1^p}e^{-ik_{e,z}z}\right)\rd k_\|.
\eeq
The solution is, simply,
\begin{multline}
\ve E_\|= 
\frac{\omega\mu j_0}{4\pi\eps_zk_0^2} \grad_\|\pd{}z\int_0^\infty \frac{k_\| J_0(k_\|\rho)}{k_{e,z}} 
\left(e^{ik_{e,z}|z|}+\frac{\hat R_0^p\hat R_1^p+\hat R_0^p}{1-\hat R_0^p\hat R_1^p}e^{ik_{e,z}z}+\frac{\hat R_0^p\hat R_1^p+\hat R_1^p}{1-\hat R_0^p\hat R_1^p}e^{-ik_{e,z}z}
-\frac{k_{e,z}}{\eps_x^{1/2}k_0}e^{i\eps_x^{1/2}k_0|z|}
\right)\rd k_\|,
\end{multline}
%
%
where the last term in brackets ensures the continuity of $\ve E_\|$ at $z=0$. The purely fluorescent decay rate is given by the rate of work done by the dipole divided by the photon energy:
\beq
\frac{-2\re\{\ve E\cdot\ve j_0^*\}}{\hbar \omega}
=\frac{\mu \eps_x|j_0|^2}{2\pi \hbar\eps_z^2k_0^2}
\re\left\{
\int_0^\infty \frac{k_\|^3}{k_{e,z}}
\left(1+\frac{2\hat R_0^p\hat R_1^p+\hat R_0^p+\hat R_1^p}{1-\hat R_0^p\hat R_1^p}\right)\rd k_\|.\right\}
\label{work1}
\eeq
The bulk value is
\beq
\Gamma_r=\frac{\mu\eps_x |j_0|^2}{2\pi \hbar\eps_z^2k_0^2}\int_0^{\eps_z^{1/2}k_0}\frac{k_\|^3}{k_{e,z}}\rd k_\|=\frac{\mu|j_0|^2k_0}{3\pi\hbar}\eps_x^{1/2}.
\eeq
Using this expression, we may rewrite the right hand side of (\ref{work1})  as
\beq
\Gamma_r\left(1+\frac{3\eps_x^{1/2}}{2\eps_z^2k_0^3}
\re\left\{
\int_0^\infty \frac{k_\|^3}{k_{e,z}}
 \frac{2\hat R_0^p\hat R_1^p+\hat R_0^p+\hat R_1^p}{1-\hat R_0^p\hat R_1^p} \rd k_\|\right\}
\right),
\eeq
from which (\ref{bperp}) follows.

\subsection{Parallel dipole}
Let now $\ve j_0=j_0\ex$. This time, both ordinary and extraordinary waves are generated.
Following the same reasoning as for the perpendicular dipole, one obtains
\begin{align}
E_z&=\frac{-\omega\mu j_0}{4\pi\eps_zk_0^2}\pd{^2}{x\dd z}\int_0^\infty\frac{J_0(k_\|\rho)}{k_{e,z}}\left(e^{ik_{e,z}|z|}+\frac{\hat R_0^p\hat R_1^p-\hat R_0^p}{1-\hat R_0^p\hat R_1^p}e^{ik_{e,z}z}+\frac{\hat R_0^p\hat R_1^p-\hat R_1^p}{1-\hat R_0^p\hat R_1^p}e^{-ik_{e,z}z}\right)k_\|\rd k_\|,\\
B_z&=\frac{-i\mu j_0}{4\pi}\pd{}{y}\int_0^\infty\frac{J_0(k_\| \rho)}{k_{o,z}}
\left(e^{ik_{0,z}|z|}+\frac{\hat R_0^s\hat R_1^s+\hat R_0^s}{1-\hat R_0^s\hat R_1^s}e^{ik_{o,z}z}+\frac{\hat R_0^s\hat R_1^s+\hat R_1^s}{1-\hat R_0^s\hat R_1^s}e^{-ik_{o,z}z}
\right)k_\|\rd k_\|.
\end{align}
With these solutions, we may write the equation for $\ve E_\|$. Noting that
\beq
\delta(\ve x)=\frac{\delta(z)}{2\pi}\int_0^\infty J_0(k_\|\rho)k_\|\rd k_\|,
\eeq
we have 
\begin{multline}
\left(\pd{^2}{z^2}+ \eps_xk_0^2\right)\ve E_\|=\frac{-i\omega\mu j_0}{2\pi}\ex \delta(z)\int_0^\infty J_0(k_\|\rho)k_\|\rd k_\|\\
\frac{-\omega\mu j_0}{4\pi\eps_zk_0^2}\grad_\|\pd{^3}{x\dd z^2}\int_0^\infty\frac{J_0(k_\|\rho)}{k_{e,z}}\left(e^{ik_{e,z}|z|}+\frac{\hat R_0^p\hat R_1^p-\hat R_0^p}{1-\hat R_0^p\hat R_1^p}e^{ik_{e,z}z}+\frac{\hat R_0^p\hat R_1^p-\hat R_1^p}{1-\hat R_0^p\hat R_1^p}e^{-ik_{e,z}z}\right)k_\|\rd k_\|\\
\frac{-\omega\mu j_0}{4\pi}\ez\times\grad_\|\pd{}{y}\int_0^\infty\frac{J_0(k_\| \rho)}{k_{o,z}}
\left(e^{ik_{0,z}|z|}+\frac{\hat R_0^s\hat R_1^s+\hat R_0^s}{1-\hat R_0^s\hat R_1^s}e^{ik_{o,z}z}+\frac{\hat R_0^s\hat R_1^s+\hat R_1^s}{1-\hat R_0^s\hat R_1^s}e^{-ik_{o,z}z}
\right)k_\|\rd k_\|.
\label{papa}
\end{multline}
Given that $(2ik)^{-1}\exp(ik|z|)$ is the solution of $\td{^2f}{z^2}+k^2f=\delta(z)$, the solution of (\ref{papa}) is
\begin{multline}
\ve E_\|=\frac{-\omega\mu j_0}{4\pi\eps_x^{1/2}k_0}\ex \delta(z)\int_0^\infty J_0(k_\|\rho)e^{i\eps_x^{1/2}k_0|z|}k_\|\rd k_\|\\
\frac{-\omega\mu j_0}{4\pi\eps_xk_0^2}\grad_\|\pd{^3}{x\dd z^2}\int_0^\infty\frac{J_0(k_\|\rho)}{k_{e,z}k_\|}\left(e^{ik_{e,z}|z|}+\frac{\hat R_0^p\hat R_1^p-\hat R_0^p}{1-\hat R_0^p\hat R_1^p}e^{ik_{e,z}z}+\frac{\hat R_0^p\hat R_1^p-\hat R_1^p}{1-\hat R_0^p\hat R_1^p}e^{-ik_{e,z}z}-\frac{k_{e,z}}{\eps_x^{1/2}k_0}e^{i\eps_x^{1/2}k_0|z|}\right)\rd k_\|\\
\frac{-\omega\mu j_0}{4\pi}\ez\times\grad_\|\pd{}{y}\int_0^\infty\frac{J_0(k_\| \rho)}{k_{o,z}k_\|}
\left(e^{ik_{0,z}|z|}+\frac{\hat R_0^s\hat R_1^s+\hat R_0^s}{1-\hat R_0^s\hat R_1^s}e^{ik_{o,z}z}+\frac{\hat R_0^s\hat R_1^s+\hat R_1^s}{1-\hat R_0^s\hat R_1^s}e^{-ik_{o,z}z}
-\frac{k_{o,z}}{\eps_x^{1/2}k_0}e^{i\eps_x^{1/2}k_0|z|}
\right)\rd k_\|,
\end{multline}
which simplifies into
\begin{multline}
\ve E_\|=
\frac{-\omega\mu j_0}{4\pi\eps_xk_0^2}\grad_\|\pd{^3}{x\dd z^2}\int_0^\infty\frac{J_0(k_\|\rho)}{k_{e,z}k_\|}\left(e^{ik_{e,z}|z|}+\frac{\hat R_0^p\hat R_1^p-\hat R_0^p}{1-\hat R_0^p\hat R_1^p}e^{ik_{e,z}z}+\frac{\hat R_0^p\hat R_1^p-\hat R_1^p}{1-\hat R_0^p\hat R_1^p}e^{-ik_{e,z}z}\right)\rd k_\|\\
\frac{-\omega\mu j_0}{4\pi}\ez\times\grad_\|\pd{}{y}\int_0^\infty\frac{J_0(k_\| \rho)}{k_{o,z}k_\|}
\left(e^{ik_{0,z}|z|}+\frac{\hat R_0^s\hat R_1^s+\hat R_0^s}{1-\hat R_0^s\hat R_1^s}e^{ik_{o,z}z}+\frac{\hat R_0^s\hat R_1^s+\hat R_1^s}{1-\hat R_0^s\hat R_1^s}e^{-ik_{o,z}z}
\right)\rd k_\|.
\end{multline}
We may now compute the fluorescent decay rate (at $z=0, \rho=0$):
\begin{multline}
\frac{-2\re\{E_x j_0^*\}}{\hbar\omega}=
\frac{\mu |j_0|^2}{4\pi\hbar} \re\left\{\int_0^\infty 
\left[
\frac{k_{e,z}}{\eps_xk_0^2}\left(1+\frac{2\hat R_0^p\hat R_1^p-\hat R_0^p-\hat R_1^p}{1-\hat R_0^p\hat R_1^p} \right)
 \right. \right.\\\left.\left.
 +  
\frac{1}{k_{o,z} }
\left(1+\frac{2\hat R_0^s\hat R_1^s+\hat R_0^s+\hat R_1^s}{1-\hat R_0^s\hat R_1^s} 
\right)\right]k_\|\rd k_\|\right\}.
\label{work2}
\end{multline}
The bulk value, $\Gamma_r$ is obtained when all reflection coefficients vanish:
\beq
\Gamma_r=\frac{\mu|j_0|^2}{4\pi\hbar}\left(
\frac{1}{\eps_xk_0^2} \int_0^{\eps_z^{1/2}k_0} k_{e,z} k_\|\rd k_\|\\
+  \int_0^{\eps_x^{1/2}k_0}\frac{1}{k_{o,z} }
 k_\|\rd k_\|\right) =
 \frac{\mu |j_0|^2k_0}{3\pi\hbar}\left(\frac{3\eps_x+\eps_z}{4\eps_x^{1/2}}\right).
 \eeq
Factorizing this expression out of (\ref{work2}), we obtain
 \beq
 \Gamma_r\left(1+\frac{3}{4k_0}\frac{4\eps_x^{1/2}}{3\eps_x+\eps_z}\re\left\{
 \int_0^\infty \left(\frac{k_{e,z}}{\eps_xk_0^2} \frac{2\hat R_0^p\hat R_1^p-\hat R_0^p-\hat R_1^p}{1-\hat R_0^p\hat R_1^p} 
+\frac1{k_{o,z}} \frac{2\hat R_0^s\hat R_1^s+\hat R_0^s+\hat R_1^s}{1-\hat R_0^s\hat R_1^s}
 \right)k_\|\rd k_\|
 \right\}\right),
 \eeq
and (\ref{bpara}) follows.

\end{document}